\documentclass[runningheads,a4paper]{llncs}

\usepackage[T1]{fontenc}
\usepackage{amsmath,amssymb}
\usepackage{graphicx}
\usepackage{booktabs}
\usepackage{array}
\usepackage{colortbl}
\definecolor{rowtint}{gray}{0.93}
\newcolumntype{R}[1]{>{\raggedright\arraybackslash}p{#1}}
\usepackage{url}
\usepackage{xspace}
\usepackage{tikz}
\usetikzlibrary{arrows.meta,positioning,shapes,fit,backgrounds,calc}
\usepackage{pgfplots}
\pgfplotsset{compat=1.17}
\usepackage[hidelinks]{hyperref}
\usepackage{algorithm}
\usepackage{algpseudocode}
\usepackage{microtype}
\newcommand{\sys}{\textsc{CG-CTI}\xspace}     

\begin{document}

\title{From Sandbox to Enforcement: Confidence-Qualified Threat Intelligence
       for Critical Infrastructure\thanks{This is a preprint of an article published in the 21st International Conference on Critical Information Infrastructures Security (CRITIS 2026).}}
\titlerunning{Confidence-Qualified CTI for Critical Infrastructure}
\author{Nikolaos Kekatos\inst{1} \and Mihaela Curc\u{a}\inst{2} \and
        Georgios Koutidis\inst{1} \and Mihai Nena\inst{2} \and
        Tom Nianios\inst{1} \and Robert-\textcommabelow{S}tefan \textcommabelow{S}andru\inst{2} \and
        Michael Ioannou\inst{3} \and Charalambos Bratsas\inst{4}}
\authorrunning{N. Kekatos et al.}
\institute{Clone Systems, Cyprus, \email{\{nkekatos,gkoutidis,tnianios\}@clone-systems.com}
   \and Romanian National Cyber Security Directorate (DNSC), Romania,
   \email{\{mihaela.curca,mihai.nena,robert.sandru\}@dnsc.ro}
   \and Bolton Technologies, Cyprus, \email{michael@bts.com.cy}
   \and International Hellenic University, Greece, \email{cbratsas@ihu.gr}}
\maketitle

\begin{abstract}
Security operations centres and national incident-response teams defending critical
infrastructure collect abundant threat data yet struggle to turn it into \emph{actionable}
intelligence. A malware sandbox produces detailed behavioural evidence, but as a large,
unranked report whose confidence is unstated. We present \sys, an operational pipeline that
converts live sandbox output (CAPEv2) into STIX~2.1, correlates it in a knowledge graph with
other critical-infrastructure sensors, and attaches to every intelligence object an explicit
\emph{confidence status} derived from provenance, cross-source corroboration, and observation
durability. This status gates automated action: only corroborated intelligence is eligible
for automated enforcement, while lower-confidence objects are routed to analyst review or
kept as context. A grounded language-model stage then narrates the
confidence-qualified evidence, where each statement either cites a supporting object or is
marked unsupported, so fabricated references are removed before analyst review. We implement \sys\ within the
CYBERGUARD project, whose consortium includes Romania's national cyber-security
directorate, and evaluate it against the live sandbox on a labelled malware
corpus, measuring conversion validity, indicator yield, technique coverage, corroboration,
enforcement eligibility, latency, and summary grounding. \sys\ turns fragmented sandbox output into corroborated,
confidence-ranked, and auditable intelligence for critical-infrastructure defence.
\end{abstract}

\section{Introduction}
\label{sec:intro}
Malware analysis is the systematic examination of malicious software to recover its
behaviour, capabilities, and indicators of compromise~\cite{ormeir2019dynamic}.
It combines \emph{static} inspection, \emph{dynamic} detonation in an instrumented sandbox,
and hybrid approaches, each with complementary blind spots: packing and obfuscation defeat
static analysis, while anti-sandbox evasion and unexecuted code paths limit dynamic
analysis~\cite{ormeir2019dynamic,dahiya2023}. Automated sandboxes
operationalise dynamic analysis at scale: Cuckoo and its successor
CAPEv2~\cite{capev2}, alongside commercial services, detonate samples and collect
API-call traces, file-system and registry changes, network traffic, dropped files, and
extracted configurations~\cite{ormeir2019dynamic}. Malware analysis is thus a mature,
well-tooled discipline; the difficulty this paper addresses is not \emph{producing}
behavioural evidence but \emph{turning it into actionable, shareable threat intelligence}
for defenders of critical infrastructure.

Security Operations Centres (SOCs) and national Computer Security Incident Response
Teams (CSIRTs) defend entities that carry security and reporting obligations under the
NIS2 Directive~\cite{nis2}. They do not suffer from a shortage of threat data; they suffer from a shortage of \emph{actionable} threat
data. A typical
security-operations team receives on the order of several thousand alerts per day and
ignores the majority of them~\cite{alahmadi2022}, and threat data is routinely fragmented across tools and
teams so that no single analyst sees the whole picture. The binding constraint is not
visibility but \emph{actionability}.

Malware sandboxes compound this problem. A dynamic-analysis system such as
CAPEv2 detonates a suspicious sample and produces detailed behavioural evidence (process
and network activity, dropped artefacts, signatures, and their ATT\&CK mapping). This is
the input a critical-infrastructure defender requires~\cite{resilagent}, yet it arrives as a large,
sandbox-specific JSON report that is technically rich but not directly consumable by a SOC
as prioritised, shareable intelligence.
Compounding this, atomic indicators of compromise (IOCs) such as file hashes, IP
addresses, and domains have a short operational lifetime, since adversaries rotate
infrastructure once indicators are burned, whereas behavioural techniques are more durable and more
expensive for an adversary to change.

This paper addresses the gap between high-fidelity sandbox output and actionable
critical-infrastructure intelligence. We present \sys, developed within the CYBERGUARD project (whose partners include the Romanian national cyber-security directorate, DNSC) and evaluated against
a live CAPEv2 sandbox, which converts the output of that sandbox into STIX~2.1, correlates
it in an OpenCTI knowledge graph with other sensors, attaches an explicit,
evidence-derived \emph{confidence status} to every intelligence object, and generates
an evidence-grounded malware summary that an analyst validates before use. Unlike prior
systems that generate cyber threat intelligence (CTI) for an analyst to read, \sys\ attaches to each object a confidence
status that gates automated enforcement.

\paragraph{Contributions.}\nopagebreak This paper makes four contributions.
\begin{itemize}
\item \sys, an operational pipeline that ingests live CAPEv2 output as STIX~2.1 and
      correlates it with other critical-infrastructure sensors in an OpenCTI knowledge graph. Unlike prior
      CAPE$\rightarrow$STIX converters, conversion is one stage of an
      evidence-to-enforcement lifecycle demonstrated up to the enforcement gate
      (Sect.~\ref{sec:arch}).
\item An \emph{operational confidence model} that maps each object, via a single total
      function over provenance, cross-source corroboration, and observation durability, to
      an ordered status (\textsf{corroborated}$\succ$\textsf{attested}$\succ$\textsf{observed})
      that \emph{gates automated enforcement}: only corroborated intelligence is
      enforce-eligible. MISP records sightings and decays indicators and STIX~2.1 carries a
      confidence property; these describe confidence, whereas we make it a control-plane
      decision that gates enforcement (Sect.~\ref{sec:confidence}).
\item \emph{Evidence-grounded malware summaries} that invert the usual
      report$\rightarrow$STIX direction of LLM-assisted threat intelligence: the model narrates
      already-converted, confidence-qualified STIX, with each statement either citing
      supporting evidence or explicitly marked unsupported; references to non-existent
      objects are dropped before delivery ($12.1\%$ at scale), so every surfaced claim is
      analyst-auditable (Sect.~\ref{sec:egmir}).
\item An evaluation on a labelled corpus detonated through the deployed sandbox,
      quantifying conversion validity, yield, coverage, confidence-based enforcement selectivity,
      latency, and the grounding of the AI summaries (Sect.~\ref{sec:eval}).
\end{itemize}

\paragraph{Novelty.} \sys\ combines in one pipeline what prior systems provide separately.
It turns confidence from an analytic score into an \emph{enforcement gate}: a per-object
status, derived only from evidence the graph already records, deciding whether an object is
auto-enforced, reviewed, or kept as context. And it inverts the direction of LLM-assisted
CTI, narrating deterministically-converted, confidence-qualified STIX rather than extracting
STIX from prose, so any statement citing evidence that does not exist is dropped before it
reaches the analyst.

\paragraph{Structure.} Section~\ref{sec:background} motivates the problem;
Sects.~\ref{sec:arch}--\ref{sec:egmir} present the architecture, the confidence model that
gates enforcement, and the evidence-grounded summaries; Sects.~\ref{sec:impl}
and~\ref{sec:eval} give the implementation and evaluation; and
Sects.~\ref{sec:related}--\ref{sec:conclusion} position the work, discuss it, state
limitations, and conclude.

\section{Background and Motivation}
\label{sec:background}
\paragraph{Malware analysis and its intelligence yield.}
Malware analysis spans static inspection (headers, strings, hashes, disassembly),
dynamic detonation in a sandbox, hybrid combinations of the two, and, increasingly,
AI-assisted reverse engineering. Static analysis is thorough but slow and defeated by
packing and obfuscation; dynamic analysis observes actual runtime behaviour and yields
immediate operational indicators (contacted hosts, dropped payloads, persistence
mechanisms), at the cost of missing dormant code paths and being susceptible to
anti-sandbox evasion~\cite{ormeir2019dynamic}. CAPEv2~\cite{capev2} is an open
dynamic-analysis system with kernel-level API monitoring, network capture, memory
forensics, and configuration extraction; its report is the primary intelligence source
\sys\ consumes.

\paragraph{CTI standards and platforms.}
Operationalising this output depends on shared representations. STIX~2.1 provides a
machine-readable object model for observables (files, addresses, domains, URLs),
higher-level constructs (malware, attack patterns, indicators), and their
relationships~\cite{stix21}; TAXII defines how such intelligence is
exchanged~\cite{taxii21}. MITRE ATT\&CK~\cite{attack} supplies a common vocabulary of
adversary tactics and techniques that lifts intelligence from atomic indicators to
durable behaviour. Threat-intelligence platforms such as MISP~\cite{misp} and
OpenCTI~\cite{opencti} store, correlate, and visualise this intelligence; \sys\ is built
on OpenCTI and represents all sensor output as STIX~2.1.

\paragraph{From data to actionable intelligence.}
Integrating threat intelligence into SOC workflows (correlation, response playbooks, threat hunting)
turns passive monitoring into intelligence-driven operations, but only if the
intelligence is prioritised and trustworthy: poorly curated feeds overload a SIEM with
false positives, and undifferentiated indicator streams worsen the very alert fatigue
they are meant to relieve. The recurring lesson is that intelligence quality, context,
and confidence matter more than raw quantity, and that behaviour-level
intelligence outlasts atomic IOCs~\cite{cti_survey}. \sys\ takes this lesson as its
design premise: it makes confidence and durability explicit, per object, so that
downstream SOC and CSIRT consumers can act on the intelligence that matters.

\section{The \sys\ Architecture}
\label{sec:arch}
This section follows evidence through the deployed pipeline: ingested by purpose-built
connectors, converted from CAPEv2 report to STIX~2.1, correlated in the OpenCTI graph, and
served to analysts through a dashboard and API.

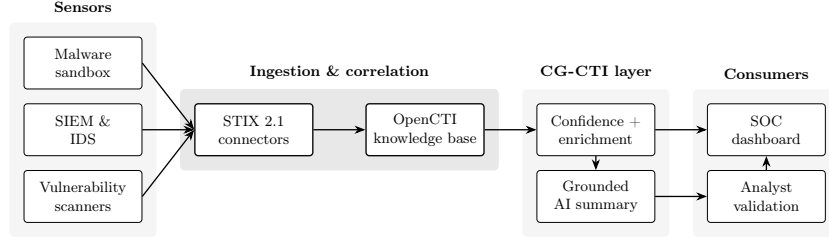
\begin{figure}[t]
\centering
\resizebox{0.90\columnwidth}{!}{%
\begin{tikzpicture}[font=\normalsize,
  box/.style={draw, fill=white, rounded corners=2pt, align=center, font=\footnotesize,
              minimum height=11mm, minimum width=25mm, inner sep=2pt},
  hub/.style={draw, fill=white, thick, rounded corners=2pt, align=center, font=\footnotesize,
              minimum height=11mm, minimum width=25mm, inner sep=2pt},
  arrow/.style={-Stealth, thick},
  band/.style={rounded corners=3pt, inner sep=3mm},
  tier/.style={font=\footnotesize\bfseries}]
\node[box] (cape) {Malware\\sandbox};
\node[box, below=3mm of cape] (det) {SIEM \&\\IDS};
\node[box, below=3mm of det]  (sc)  {Vulnerability\\scanners};
\node[hub, right=11mm of det]  (conn) {STIX~2.1\\connectors};
\node[hub, right=11mm of conn] (octi) {OpenCTI\\knowledge base};
\node[box, right=11mm of octi] (conf)  {Confidence +\\enrichment};
\node[box, below=3mm of conf]  (egmir) {Grounded\\AI summary};
\node[box, right=11mm of conf] (dash) {SOC\\dashboard};
\node[box, below=3mm of dash]  (val)  {Analyst\\validation};
\draw[arrow] (cape.east) -- (conn.west);
\draw[arrow] (det.east)  -- (conn.west);
\draw[arrow] (sc.east)   -- (conn.west);
\draw[arrow] (conn) -- (octi);
\draw[arrow] (octi.east)  -- (conf.west);
\draw[arrow] (conf.south) -- (egmir.north);
\draw[arrow] (conf.east)  -- (dash.west);
\draw[arrow] (egmir.east) -- (val.west);
\draw[arrow] (val.north)  -- (dash.south);
\begin{scope}[on background layer]
  \node[band, fill=black!4, fit=(cape)(sc)]    (bsrc)   {};
  \node[band, fill=black!9, fit=(conn)(octi)]  (bcore)  {};
  \node[band, fill=black!4, fit=(conf)(egmir)] (blayer) {};
  \node[band, fill=black!4, fit=(dash)(val)]   (bcons)  {};
\end{scope}
\node[tier, above=1mm of bsrc.north]   {Sensors};
\node[tier, above=1mm of bcore.north]  {Ingestion \& correlation};
\node[tier, above=1mm of blayer.north] {\sys\ layer};
\node[tier, above=1mm of bcons.north]  {Consumers};
\end{tikzpicture}}
\Description{Block diagram of the CG-CTI pipeline in four bands read left to right. Sensors: malware sandbox, SIEM and IDS, vulnerability scanners. Ingestion and correlation: STIX 2.1 connectors feeding an OpenCTI knowledge base. CG-CTI layer: a confidence and enrichment stage above a grounded AI summary stage. Consumers: a SOC dashboard fed by the confidence stage, and analyst validation fed by the AI summary and reporting back into the dashboard.}%
\caption{The \sys\ pipeline. STIX~2.1 connectors normalise heterogeneous sensor output
into one OpenCTI knowledge graph, where the \sys\ layer qualifies each object's
confidence, corroborates it against lightweight open-feed lookups, and generates an
evidence-grounded AI summary that an analyst validates before it reaches the
SOC dashboard. The sensors are a malware sandbox (CAPEv2), SIEM and host/network IDS,
and vulnerability scanners; the SOC dashboard is the operator console.}
\label{fig:arch}
\end{figure}

\paragraph{Ingestion: seven connectors into one knowledge base.}
\sys\ is built on OpenCTI and fed by seven purpose-built STIX~2.1 connectors that group
into three roles: a \emph{malware-analysis} tier (the sandbox connector), a
\emph{detection} tier (host and network intrusion detection), and an
\emph{exposure-assessment} tier (vulnerability and code scanners). Each connector \emph{normalises} before publishing: scanner findings arrive with their own
identifier and severity schemes, and are mapped onto CVE/CWE and a common scale, so
heterogeneous output lands in one correlatable graph (Fig.~\ref{fig:arch}). The malware and detection tiers are this
paper's focus, since they carry the behavioural evidence a critical-infrastructure SOC
acts on; the assessment tier situates that evidence within the broader attack-surface
context and shares the same knowledge base, so no bespoke per-consumer integration is
needed. The specific products behind each connector are listed in Sect.~\ref{sec:impl}.

\paragraph{Malware path: sandbox report to STIX.}
The sandbox connector retrieves each completed CAPEv2 report and normalises it into
STIX~2.1 objects: the sample file and its hashes, a malware object typed from the reported
family, the network indicators (contacted addresses, domains, and URLs) reconstructed from
the sandbox telemetry, and attack-pattern objects mapping sandbox signatures to MITRE
ATT\&CK techniques. Relationships tie these to a source identity so provenance is preserved
in the graph. The concrete STIX types and the report fields they draw from are given in
Sect.~\ref{sec:impl}.

\begin{figure}[t]
\centering
\definecolor{navy}{RGB}{31,41,55}
\definecolor{headerbg}{RGB}{234,237,241}
\definecolor{accent}{RGB}{37,99,235}
\definecolor{appbg}{RGB}{245,246,248}
\definecolor{cardbd}{RGB}{221,224,230}
\definecolor{ink}{RGB}{31,41,55}
\definecolor{muted}{RGB}{120,130,145}
\definecolor{green}{RGB}{22,143,74}
\definecolor{greent}{RGB}{222,242,229}
\definecolor{amber}{RGB}{176,124,20}
\definecolor{ambert}{RGB}{250,240,213}
\definecolor{grayt}{RGB}{235,237,240}
\resizebox{0.78\textwidth}{!}{%
\begin{tikzpicture}[font=\fontfamily{phv}\selectfont, x=1cm, y=1cm]
\fill[appbg] (0,0) rectangle (16,9);
\fill[headerbg] (0,8.05) rectangle (16,9);
\fill[black!14] (0,8.02) rectangle (16,8.05);              
\fill[white] (0,0) rectangle (3.15,8.02);                  
\fill[cardbd] (3.13,0) rectangle (3.15,8.02);              
\draw[cardbd, line width=0.7pt] (0,0) rectangle (16,9);    
\node[anchor=west] at (0.30,8.55) {\includegraphics[height=0.42cm]{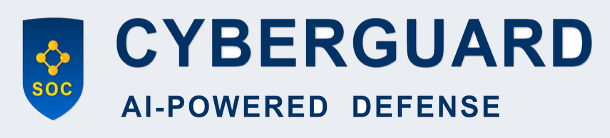}};
\draw[cardbd, line width=0.5pt] (2.60,8.30) -- (2.60,8.80);
\node[anchor=west, text=ink, font=\scriptsize] at (2.76,8.55) {Threat Intelligence Console};
\draw[cardbd, line width=0.5pt] (13.60,8.28) -- (13.60,8.82);
\node[anchor=west, text=muted, font=\scriptsize] at (13.76,8.55) {Last sync 12:04};
\fill[accent!9, rounded corners=1pt] (0.10,7.30) rectangle (3.02,7.78);
\fill[accent] (0.10,7.30) rectangle (0.17,7.78);
\foreach \y/\ic/\tc/\wt/\lab in {%
  7.54/accent/accent/\bfseries/Overview,
  7.00/black!22/ink/\mdseries/Malware,
  6.46/black!22/ink/\mdseries/SIEM \& IDS,
  5.92/black!22/ink/\mdseries/Scanners,
  5.38/black!22/ink/\mdseries/ATT\&CK,
  4.84/black!22/ink/\mdseries/Sharing}{
  \fill[\ic, rounded corners=1.5pt] (0.30,\y-0.10) rectangle (0.52,\y+0.10);
  \node[anchor=west, text=\tc, font=\scriptsize\wt] at (0.66,\y) {\lab};
}
\node[anchor=west, text=ink, font=\bfseries] at (3.45,7.62) {Security overview};
\draw[fill=white, draw=cardbd, line width=0.5pt, rounded corners=3pt] (14.52,7.44) rectangle (15.70,7.78);
\node[text=muted, font=\scriptsize] at (15.11,7.61) {Last 24\,h};
\foreach \x/\lab/\val/\vc/\dc in {%
  3.45/STIX OBJECTS/489/ink/muted,
  6.53/CORROBORATED/48/green/green,
  9.61/ATT\&CK/39/ink/accent,
  12.69/SAMPLES/35/ink/muted}{
  \draw[fill=white, draw=cardbd, line width=0.5pt, rounded corners=3pt] (\x,6.25) rectangle (\x+2.95,7.20);
  \node[anchor=west, text=muted, font=\tiny] at (\x+0.20,6.98) {\lab};
  \node[anchor=west, text=\vc, font=\large\bfseries] at (\x+0.20,6.52) {\val};
  \fill[\dc] (\x+2.70,6.98) circle (0.07);
}
\fill[greent, rounded corners=2pt] (7.92,6.35) rectangle (9.40,6.66);
\node[text=green, font=\tiny] at (8.66,6.505) {enforce-eligible};
\draw[fill=white, draw=cardbd, line width=0.5pt, rounded corners=3pt] (3.45,0.45) rectangle (9.45,6.00);
\node[anchor=west, text=ink, font=\small\bfseries] at (3.65,5.66) {Malware families};
\node[anchor=east, text=muted, font=\scriptsize] at (9.26,5.66) {35 samples};
\fill[cardbd] (3.60,5.44) rectangle (9.30,5.455);
\foreach \y/\fam/\tc/\tint/\dc/\tx/\lab in {%
  5.14/njRAT/15/greent/green/green/corroborated,
  4.56/AgentTesla/4/greent/green/green/corroborated,
  3.98/RemcosRAT/3/ambert/amber/amber/attested,
  3.40/AsyncRAT/10/ambert/amber/amber/attested,
  2.82/WannaCry/1/grayt/muted/muted/observed,
  2.24/CoinMiner/5/grayt/muted/muted/observed,
  1.66/Formbook/3/grayt/muted/muted/observed}{
  \node[anchor=west, text=ink, font=\scriptsize] at (3.68,\y) {\fam};
  \node[anchor=west, text=muted, font=\scriptsize] at (6.05,\y) {\tc\,tech};
  \fill[\tint, rounded corners=3pt] (7.35,\y-0.16) rectangle (9.22,\y+0.16);
  \fill[\dc] (7.56,\y) circle (0.05);
  \node[text=\tx, font=\tiny] at (8.44,\y) {\lab};
}
\draw[fill=white, draw=cardbd, line width=0.5pt, rounded corners=3pt] (9.58,0.45) rectangle (15.70,6.00);
\node[anchor=west, text=ink, font=\small\bfseries] at (9.78,5.66) {Confidence status};
\node[anchor=east, text=muted, font=\scriptsize] at (15.52,5.66) {489 objects};
\fill[cardbd] (9.73,5.44) rectangle (15.55,5.455);
\foreach \y/\lab/\w/\col/\n in {%
  5.06/Corroborated/0.69/green/48,
  4.31/Attested/3.17/amber/222,
  3.56/Observed/3.13/black!28/219}{
  \node[anchor=west, text=ink, font=\scriptsize] at (9.78,\y) {\lab};
  \fill[grayt, rounded corners=2pt] (11.85,\y-0.13) rectangle (15.02,\y+0.13);
  \fill[\col, rounded corners=2pt] (11.85,\y-0.13) rectangle (11.85+\w,\y+0.13);
  \node[anchor=west, text=muted, font=\scriptsize] at (15.14,\y) {\n};
}
\fill[cardbd] (9.73,3.06) rectangle (15.55,3.075);
\foreach \y/\dc/\txt in {%
  2.68/green/{Only \textbf{corroborated} is enforce-eligible (9.8\%)},
  2.10/accent/{Grounded AI summaries: every claim cited or flagged},
  1.52/amber/{Latency p50/p95: 157 / 194\,s},
  0.94/muted/{Enrichment: MalwareBazaar + ThreatFox}}{
  \fill[\dc] (9.92,\y) circle (0.055);
  \node[anchor=west, text=ink, font=\tiny] at (10.08,\y) {\txt};
}
\end{tikzpicture}}
\Description{Mock-up of the operator console drawn as a web dashboard. A grey header shows the CYBERGUARD wordmark and the title Threat Intelligence Console. A left sidebar lists Overview (selected), Malware, SIEM and IDS, Scanners, ATT&CK and Sharing. The main area, titled Security overview for the last 24 hours, shows four indicator cards: 489 STIX objects, 48 corroborated with an enforce-eligible badge, 39 ATT&CK techniques and 35 samples. Below, a Malware families panel lists seven families (njRAT, AgentTesla, RemcosRAT, AsyncRAT, WannaCry, CoinMiner, Formbook) with technique counts and a status tag of corroborated, attested or observed, and a Confidence status panel shows horizontal bars for Corroborated 48, Attested 222 and Observed 219 objects, with a note that only corroborated objects are enforce-eligible (9.8 percent).}%
\caption{The \sys\ operator console in the CYBERGUARD testbed. It is drawn, not
screenshotted, so it stays legible in print. All values are measured on the 35-sample corpus of
Sect.~\ref{sec:eval}. Each object carries a confidence status, and only
\textsf{corroborated} objects are enforce-eligible.}
\label{fig:dashboard}
\end{figure}
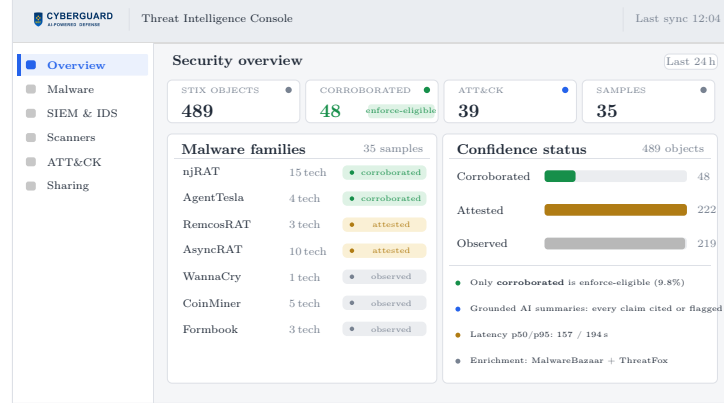

\paragraph{Consumption: dashboard and API.}
A REST service exposes the knowledge base to a SOC-facing dashboard through endpoints
for statistics, timelines, per-source alerts, malware families and their IOCs, and
ATT\&CK-technique coverage, so an analyst can move from a high-level indicator down to
the underlying sandbox evidence within one consistent data model (Fig.~\ref{fig:dashboard}).
The current dashboard service is a pilot-grade read layer, and we return to its hardening in
Sect.~\ref{sec:limits}.

\paragraph{Correlation in the OpenCTI graph.}
Because every connector writes into one OpenCTI knowledge base, \sys\ reuses OpenCTI's
\emph{native} graph rather than building a separate store: identical observables emitted
by different sensors (a hash seen by the sandbox and flagged by a detection sensor, a
domain in both sandbox telemetry and a network alert) are merged into a single node, and
four relationship types connect analyses into a navigable graph:
\textsf{drops} (malware$\rightarrow$file), \textsf{uses}
(malware$\rightarrow$attack-pattern) and \textsf{communicates-with}
(malware$\rightarrow$address, domain or URL) from the sandbox path, and
\textsf{indicates} (indicator$\rightarrow$object) from the detection and assessment tiers. This native correlation is the signal the confidence model's
corroboration axis reads (Sect.~\ref{sec:confidence}); the graph is not only a visualisation; it supplies the cross-source evidence on which confidence is
computed.

\section{An Operational Confidence Model: From Evidence to Enforcement}
\label{sec:confidence}
Raw sandbox and feed output arrives with heterogeneous scoring conventions: the sandbox
emits a malware score, the vulnerability scanner a CVSS-derived \emph{severity} score, and
most connectors no comparable score at all. \sys\ replaces these scattered
heuristics with one model attaching an explicit confidence status to every object.

\paragraph{Three evidence axes.}
\sys\ derives, for every intelligence object $o$, a confidence status from three
axes the knowledge base already records.
\emph{(i)~Provenance} $p(o)$ is a per-source base weight reflecting how directly the
source witnessed the artefact: a dynamic behavioural observation from the sandbox
(the sample actually contacted a host, dropped a file, or exhibited a signature) ranks
above a static pattern match, which ranks above an external feed lookup. The ad-hoc
per-connector scores already in the deployment are special cases of $p$: the sandbox's
malware score and the vulnerability connector's CVSS-derived score
($\mathit{conf}=\min(10\cdot\mathrm{CVSS},100)$) are normalised into the provenance
axis $p$, which \sys\ unifies onto one scale. We use $p$ as an ordinal rank,
not a numeric threshold: \emph{high} for a direct dynamic behavioural observation (the
sample \emph{exhibited} the behaviour) and \emph{low} for a passive atomic observation
(an address merely contacted, a hash of a dropped file).
\emph{(ii)~Corroboration} $k(o)$ counts the sources beyond the originating sandbox that
attest the same observable, another sensor already in the graph or an external-feed hit,
so $k(o){=}0$ means only the sandbox saw it. Because every
connector writes into one OpenCTI graph, an identical hash, domain, or technique seen by
the sandbox and a second source is automatically correlated.
\emph{(iii)~Durability} $d(o)$ is the class of the observable: an \emph{atomic} IOC
(hash, IP, domain, URL), which adversaries rotate quickly, versus a \emph{behavioural}
object (a signature or ATT\&CK technique), which is more durable. This axis
operationalises the CTI survey's~\cite{cti_survey} finding that behavioural techniques
outlive atomic indicators, directly in the confidence
model.

\paragraph{Status and rule.}
The three axes map to an ordered confidence status
$C(o)\in\{\textsf{corroborated}\succ\textsf{attested}\succ\textsf{observed}\}$ by the rule
\[
C(o)=
\begin{cases}
\textsf{corroborated} & k(o)\ge 1\ \text{(at least one source beyond the sandbox)},\\
\textsf{attested}     & k(o)=0\ \text{and}\ \big(d(o)=\text{behavioural}\ \text{or}\ p(o)=\text{high}\big),\\
\textsf{observed}     & k(o)=0\ \text{and}\ d(o)=\text{atomic}\ \text{and}\ p(o)=\text{low}.
\end{cases}
\]
Durability affects both initial qualification and subsequent ageing: behavioural evidence
can remain \textsf{attested} without corroboration and retains that status, whereas an
atomic IOC that is never corroborated ages out of the actionable set, so the model prefers
durable evidence exactly as the survey recommends.
Table~\ref{tab:conf} states what a SOC does with each level; only
\textsf{corroborated} intelligence is eligible for automated enforcement, keeping
low-confidence indicators out of the automated-enforcement path. The
thresholds above are design defaults to be tuned against operator feedback
(Sect.~\ref{sec:limits}); the contribution is the explicit, per-object,
evidence-derived status, not the specific cut points.

\paragraph{From evidence to enforcement.}
The model is a three-stage chain: the evidence axes form a per-object trust assessment, the
rule maps that assessment to an ordered \emph{confidence status}, and the status gates a
\emph{SOC action} (Table~\ref{tab:conf}). Two properties keep the mapping principled rather
than ad hoc. It is a total function: every object receives a status and none is left
unranked. It is also monotone in each axis, since raising an object's corroboration,
durability, or provenance never lowers its status; the status is therefore an
order-preserving image of the evidence, not a lookup table. Each cut point follows a stated
principle: attestation by a corroborating source beyond the originating sandbox ($k\ge1$) is the weakest evidence that licenses automated
action, because a single sensor can be evaded; a durable behavioural technique or a directly
witnessed behaviour merits analyst review even uncorroborated; and a lone passive atomic
indicator is kept as context but never enforced.

\paragraph{Lightweight corroborating enrichment.}
The corroboration axis $k(o)$ is strengthened by a deliberately minimal
enrichment step, kept small so its effect stays measurable. For each sandbox
observable \sys\ issues at most two open-feed lookups, a file-hash lookup against
MalwareBazaar (family, tags) and an indicator lookup against ThreatFox (threat
label)~\cite{abusech}, reusing a single free abuse.ch key. A feed hit on the observable, a
hash present in MalwareBazaar or an indicator listed in ThreatFox, counts as an external-feed
attestation and promotes the object toward \textsf{corroborated}; the feed's family or
threat tags are recorded, but exact-label agreement is not required for promotion, because
community labels are noisy. This keeps enrichment cheap and its effect
measurable. One caveat matters for reading the results, and we quantify it in
Sect.~\ref{sec:eval}: for a MalwareBazaar-drawn corpus a hash hit corroborates almost by
construction, so the genuinely independent signals are ThreatFox network-indicator hits and
cross-sensor corroboration.

\begin{table}[t]
\centering
\caption{Confidence status, the evidence that yields it, and the SOC action it gates.}
\label{tab:conf}
\footnotesize
\setlength{\tabcolsep}{5pt}
\begin{tabular}{@{}R{2.6cm}R{4.6cm}R{3.8cm}@{}}
\toprule
\textbf{Status} & \textbf{Evidence} & \textbf{SOC action} \\
\midrule
\textsf{corroborated} & a source beyond the sandbox attests it ($k{\ge}1$) & enforce-eligible; shareable \\
\textsf{attested}     & sandbox-only ($k{=}0$), behavioural or high-provenance & analyst-review queue \\
\textsf{observed}     & sandbox-only ($k{=}0$), low-provenance atomic IOC & context only, not enforced \\
\bottomrule
\end{tabular}
\end{table}

Qualification is applied to every object as it enters the graph: the corroborating set $K$
collects sources beyond the originating sandbox, an open-feed lookup may add to it for
atomic observables, and the status follows Table~\ref{tab:conf}.

\section{Evidence-Grounded AI Malware Summaries}
\label{sec:egmir}
The confidence-qualified graph is machine-actionable, but an analyst or a CSIRT duty
officer still needs a readable account of \emph{what a sample does and why it matters}.
Producing that by hand is slow; producing it with a large language model (LLM) applied
to raw sandbox logs is fast but untrustworthy, because the model can assert behaviours
the evidence does not support. \sys\ takes a middle path: an \emph{Evidence-Grounded
Malware Summary} in which supported statements are tied to the STIX
objects and sandbox evidence that justify them.

\paragraph{Grounding by construction.}
The AI stage never sees the raw CAPEv2 log. Instead, \sys\ selects a compact evidence
set from the confidence-qualified STIX graph, the \texttt{File}, \texttt{Malware},
network observables, and ATT\&CK \texttt{AttackPattern}s for the sample, each associated
with its derived confidence status, and passes only that structured evidence to the model. The
model is instructed to write an executive summary, observed behaviours, ATT\&CK
techniques, IOCs, likely impact, and recommended mitigations, and to \emph{cite, for
each statement, the identifier(s) of the supporting STIX object(s)}. A statement with no
supporting object is shown but marked unsupported, so the claim is visible as such instead
of reading as fact; only a statement citing an identifier that does not exist is dropped. Because the evidence is already confidence-qualified,
the summary inherits the same qualification: a behaviour asserted only from
\textsf{observed} (uncorroborated) evidence is marked accordingly. Algorithm~\ref{alg:egmir} makes the
grounding check explicit: every generated statement is scored against the evidence set,
and the report's \emph{grounding rate} (Q7) is the fraction of statements traceable to a
valid evidence identifier. In the algorithm, \textsc{SelectEvidence} assembles the evidence
set $E$ from the graph, \textsc{Prompt} renders it for the model, \textsc{Cites} returns the
evidence identifiers a statement references, and \textsc{AnalystValidate} is the human
approval gate.

\begin{algorithm}[t]
\caption{Evidence-grounded malware summary and grounding check.}
\label{alg:egmir}
\footnotesize
\begin{algorithmic}[1]
\Require sample $s$; confidence-qualified STIX graph $G$; self-hosted LLM $M$
\Ensure report $R$; grounding metrics
\State $E \gets \textsc{SelectEvidence}(G, s)$ \Comment{evidence objects with associated $C(\cdot)$}
\State assign stable ids $e_1,\dots,e_n$ to $E$; $\mathit{ids} \gets \{e_1,\dots,e_n\}$
\State $R \gets M(\textsc{Prompt}(E))$ \Comment{JSON: statements, each with cited evidence ids}
\State $g \gets f \gets u \gets 0$
\ForAll{statement $t \in R$}
  \If{$\textsc{Cites}(t) = \emptyset$} \State $f \gets f+1$ \Comment{model-flagged unsupported}
  \ElsIf{$\textsc{Cites}(t) \subseteq \mathit{ids}$} \State $g \gets g+1$ \Comment{grounded}
  \Else\ \State $u \gets u+1$; drop $t$ \Comment{cites unknown id: hallucinated ref}
  \EndIf
\EndFor
\State $R \gets \textsc{AnalystValidate}(R)$; write $R$ and status back to $G$
\State \Return $R,\ (g, f, u)$
\end{algorithmic}
\end{algorithm}

\paragraph{Analyst validation.}
The generated report is a draft for analyst review, not an autonomous verdict. An analyst validates,
edits, or rejects it in the dashboard; the validated report and its analyst status are
written back to the graph as provenance, so downstream consumers see human-approved
intelligence. This human-in-the-loop design follows the CTI survey's~\cite{cti_survey} central caution, that AI accelerates but cannot replace analyst judgement, and makes the AI output
auditable: each claim can be traced to evidence and each report to an approver. We
evaluate the summaries for evidence-reference grounding (Sect.~\ref{sec:eval}).

\section{Implementation}
\label{sec:impl}
\sys\ is deployed with Docker Compose. The knowledge base is OpenCTI~6.3.9 backed by
Elasticsearch~8.15.3, Redis, RabbitMQ, and MinIO, with two ingestion workers. The seven
connectors are Python modules using the OpenCTI client library and the \texttt{stix2}
toolkit; each polls its source, builds a STIX~2.1 bundle, and publishes it to the
platform. Conversion uses only standard STIX~2.1 types --- \texttt{Bundle},
\texttt{Identity}, \texttt{Indicator}, \texttt{ObservedData}, \texttt{Sighting},
\texttt{Relationship}, \texttt{AttackPattern}, \texttt{Vulnerability}, \texttt{Malware}
and the file, address, domain, URL and network-traffic observables --- with no custom
(\texttt{x\_}) properties and no extension definition. Observables are wrapped in
\texttt{Indicator} and \texttt{ObservedData} objects, and the standard numeric
\texttt{confidence} property is carried
on those (and on \texttt{AttackPattern}/\texttt{Vulnerability}), never on the
Cyber-observable Objects themselves, which STIX~2.1 gives no \texttt{confidence}
property. The three statuses are not persisted as a STIX label: the platform stores the
standard numeric \texttt{confidence} per source, and the categorical status is derived on
read from the evidence the graph already records, so it is recomputed as evidence
accumulates instead of being fixed at ingest. The tiers differ in how much of that vocabulary they need: detonation yields a
\emph{subject}, the artefacts it touched and the techniques it exhibited, so the sandbox
connector emits the full set including \texttt{Malware}, whereas the detection and
assessment connectors are \texttt{Indicator}-centred, a rule firing having no subject to
attribute behaviour to. The SOC-facing read layer is a FastAPI service over the platform's GraphQL
API, and the dashboard is a React/Vite single-page application. The malware source is
the CYBERGUARD live CAPEv2~(v2.4) sandbox, which the sandbox connector
reaches over CAPE's task-report API. The connector maps each report to STIX~2.1 as a
\texttt{File} observable with MD5/SHA-1/SHA-256; a \texttt{Malware} object typed from the
reported family; \texttt{NetworkTraffic} with \texttt{IPv4Address}, \texttt{DomainName}, and
\texttt{URL} observables from the DNS/TCP/UDP/HTTP telemetry; and \texttt{AttackPattern}
objects from the sandbox's ATT\&CK mapping. The detection and assessment tiers are host and
network intrusion-detection sensors (Wazuh/OpenSearch and Suricata) and vulnerability and
code scanners (Nmap, OWASP~ZAP, Semgrep, OSV-Scanner). A separate REST wrapper around CAPE exists in the
codebase but is a development scaffold and is \emph{not} on the measured path; \sys\
ingests directly from the sandbox's report interface. The summarisation stage
(Sect.~\ref{sec:egmir}) is served by a \emph{self-hosted} Qwen2.5-14B~\cite{qwen25}
model behind an OpenAI-compatible endpoint (vLLM~\cite{vllm}) on operator
infrastructure, so no sample, report, or generated summary leaves the organisation. This is
a deliberate data-sovereignty choice for operators and national CSIRTs that cannot send
malware evidence to a third-party cloud. Enrichment is the one exception: it discloses
hashes and network indicators to abuse.ch, which an operator can disable at the cost of
the corroboration axis.

\section{Evaluation}
\label{sec:eval}
This section evaluates \sys\ on a labelled malware corpus detonated through the
deployed CAPEv2 sandbox and ingested end-to-end into the knowledge base.

\paragraph{Corpus and method.}
We assemble a labelled corpus of $35$ recent malware samples across seven families
(WannaCry and CoinMiner $\times9$, njRAT $\times5$, AsyncRAT $\times4$, AgentTesla and
RemcosRAT $\times3$, Formbook $\times2$), drawn from MalwareBazaar~\cite{abusech} with
community family labels, and detonate each through the deployed CAPEv2 sandbox under a
Windows execution profile with network routing enabled. The per-family counts reflect the
most recent samples available per family on MalwareBazaar at collection time; WannaCry and
CoinMiner together account for $18$ of the $35$ and are among the lowest-yield families,
which lowers the reported medians. For every sample
we retrieve the sandbox report through CAPE's task-report API and run it through the
full \sys\ pipeline into OpenCTI, so the measurements exercise the deployed
system end-to-end. We answer seven questions:
(Q1)~\emph{conversion validity}, what fraction of reports convert to schema-valid
STIX~2.1 \emph{objects} --- the \texttt{File}, network observables, and ATT\&CK
\texttt{AttackPattern}s the connector emits; the derived \texttt{Malware} object is the hub of the
malware path, the source of every \textsf{drops}, \textsf{uses} and
\textsf{communicates-with} edge. It anchors the graph but is not itself evidence, so it is
excluded from the evaluated count: the $489$ counts evidence objects, not the whole graph, which
also holds the $35$ \texttt{Malware} anchors and their relationships; (Q2)~\emph{yield}, how many network and file observables (contacted hosts, domains, URLs,
the file hash) become shareable indicators; registry modifications are behavioural volume
folded into the \texttt{Malware} object and are excluded, as are ATT\&CK techniques, which
become objects but are not raw observables.
(Q3)~\emph{ATT\&CK coverage}, how many distinct techniques are recovered from the
sandbox \texttt{ttps} mapping; (Q4)~\emph{corroboration}, how often the \emph{atomic}
observables the confidence model can look up (file hash, IPs, domains) are attested by an
open feed; (Q5)~\emph{confidence effect}, the confidence-status distribution over all
evaluated evidence objects and the fraction eligible for automated enforcement; (Q6)~\emph{latency}, submission-to-STIX end-to-end
time (p50/p95); and (Q7)~\emph{summary grounding}, what fraction of AI-summary statements
are traceable to a supporting STIX object, and the rate of flagged/unsupported statements,
with a light analyst check. Repository community labels are used as reference family
labels; ATT\&CK coverage uses the sandbox's own technique mapping and is reported as
recovered coverage, not detection accuracy.

\paragraph{Results.}
Table~\ref{tab:eval} reports the measured results. \emph{Conversion (Q1):} all $35$
reports converted end-to-end to schema-valid STIX~2.1, $489$ objects across the corpus,
$100\%$ structurally valid across the evaluated file, network, and attack-pattern objects.
\emph{Yield (Q2):} median $5$ per sample (mean $7.6$, range $1$--$23$). Registry
modifications, which reach $1339$ for one RAT sample (and exceed $100$ for only four), are
excluded as behavioural volume. Yield varies
with family: interactive RATs (njRAT, AsyncRAT, RemcosRAT) surface $15$--$16$ indicators at
the median while WannaCry and Formbook surface $1$--$3$, representative behaviour that did
not fully trigger under the default profile, a coverage or execution-profile effect that is itself a
signal (a low-yield sample is flagged as such, not silently under-reported)
and that motivates the sandbox-hardening and multi-run directions of Sect.~\ref{sec:limits}. \emph{ATT\&CK coverage (Q3):} $39$ distinct
techniques are recovered across the corpus, mean $6.3$ per sample. \emph{Latency (Q6):}
end-to-end submission-to-STIX time is dominated by sandbox detonation (analysis
$p50{=}157$\,s, $p95{=}194$\,s); the CAPE$\rightarrow$STIX transform is sub-millisecond, so
the intelligence is available as soon as the sandbox finishes. \emph{Corroboration (Q4):}
$48$ of $168$ atomic observables ($28.6\%$) are attested by an open feed: $34$ MalwareBazaar
hash hits and $14$ ThreatFox \emph{network}-indicator hits. For a MalwareBazaar-drawn
corpus a file's \emph{presence} there is near-guaranteed ($34$ of $35$ hashes), so hash
attestation is a consistency check, not independent evidence; only the $14$ ThreatFox hits
on command-and-control infrastructure --- not the corpus-selection key --- are genuinely
independent here. Community \emph{family labels} agree with the sandbox in only $30.8\%$ of
labelled cases (an AsyncRAT tagged \emph{XWorm}), so promotion uses feed \emph{presence and
tags}, not exact family match.
\emph{Confidence effect (Q5):} of $489$ objects, $48$ ($9.8\%$) are corroborated and hence
enforce-eligible; the rest go to analyst review (behavioural \textsf{attested}) or context
(uncorroborated \textsf{observed}). We do not claim the other $441$ are false positives,
only that they lack the corroboration the gate requires.
Because $34$ of the $48$ are hash hits near-guaranteed for a MalwareBazaar-sourced corpus,
the $9.8\%$ is corpus-dependent.
\emph{Summary grounding (Q7):} of $437$ statements, $121$ ($27.7\%$) cite an object, $263$
($60.2\%$) carry no citation and are surfaced as explicitly unsupported, and $53$ ($12.1\%$)
cite a non-existent object and are \emph{dropped} (Algorithm~\ref{alg:egmir}). So
\emph{$0$ hallucinated references reach the analyst}, and nothing flagged is silently
discarded. Both classes concentrate on the $24$ lowest-yield samples; the $11$
highest-yield ground $79\%$, so the aggregate reflects corpus skew, not model degradation.

\begin{table}[t]
\centering
\caption{Evaluation results on the 35-sample corpus, measured against the live CAPEv2
sandbox and the self-hosted Qwen2.5-14B LLM.}
\label{tab:eval}
\footnotesize
\begin{tabular}{@{}R{4.3cm}R{2.6cm}R{4.3cm}@{}}
\toprule
\textbf{Question} & \textbf{Metric} & \textbf{Result} \\
\midrule
Q1 Conversion validity & \% valid & $\mathbf{100\%}$ ($489/489$ objects, $35/35$ reports) \\
Q2 Yield (network+file) & per sample & median $\mathbf{5}$ (mean $7.6$, range $1$--$23$) \\
Q3 ATT\&CK coverage & techniques & $\mathbf{39}$ unique (mean $6.3$/sample) \\
Q4 Corroboration & \% corroborated & $\mathbf{28.6\%}$ ($48/168$: $34$ hash, $\mathbf{14}$ net.) \\
Q5 Confidence effect & enforce-eligible & $\mathbf{48/489}$ objects ($9.8\%$); rest to review/context \\
Q6 Latency (submit$\to$STIX) & s (p50/p95) & $\mathbf{157}$ / $\mathbf{194}$ (transform $<1$\,ms) \\
Q7 Summary grounding & grnd/flag/dropped & $\mathbf{27.7}$ / $60.2$ / $12.1$\,\% ($\mathbf{0}$ delivered) \\
\bottomrule
\end{tabular}
\end{table}

\begin{figure}[t]
\centering
\begin{tikzpicture}
\begin{axis}[
  ybar, width=0.92\textwidth, height=3.4cm, ymode=log, log origin=infty,
  ymin=1, ymax=100, ylabel={count (log)}, ylabel style={font=\footnotesize},
  bar width=6pt, enlarge x limits=0.09,
  symbolic x coords={njRAT,AgentTesla,RemcosRAT,AsyncRAT,WannaCry,CoinMiner,Formbook},
  xtick=data, xticklabel style={font=\scriptsize, rotate=20, anchor=east},
  ytick={1,10,100}, yticklabel style={font=\scriptsize},
  legend style={font=\scriptsize, at={(0.5,1.02)}, anchor=south, legend columns=2, draw=none},
  nodes near coords, nodes near coords style={font=\tiny}, point meta=explicit symbolic]
\addplot+[fill=black!18, draw=black!55] coordinates {
  (njRAT,15)[15] (AgentTesla,4)[4] (RemcosRAT,16)[16] (AsyncRAT,15)[15]
  (WannaCry,3)[3] (CoinMiner,5)[5] (Formbook,1)[1]};
\addplot+[fill=black!45, draw=black!70] coordinates {
  (njRAT,15)[15] (AgentTesla,4)[4] (RemcosRAT,3)[3] (AsyncRAT,10)[10]
  (WannaCry,1)[1] (CoinMiner,5)[5] (Formbook,3)[3]};
\legend{Network + file observables (median yield), ATT\&CK techniques (median)}
\end{axis}
\end{tikzpicture}
\Description{Grouped bar chart on a logarithmic count axis from 1 to 100 for seven malware families: njRAT, AgentTesla, RemcosRAT, AsyncRAT, WannaCry, CoinMiner and Formbook. Light bars give the median network plus file observable yield and dark bars the median number of ATT&CK techniques, labelled with their values: njRAT 15 and 15, AgentTesla 4 and 4, RemcosRAT 16 and 3, AsyncRAT 15 and 10, WannaCry 3 and 1, CoinMiner 5 and 5, Formbook 1 and 3.}%
\caption{Median per-family network+file yield and technique coverage across the 35-sample
corpus (log scale). Interactive RATs (njRAT, AsyncRAT, RemcosRAT) surface the most shareable
indicators, whereas WannaCry and Formbook surface few under the default profile, a low-yield
signal \sys\ reports explicitly (Sect.~\ref{sec:limits}).}
\label{fig:yield}
\end{figure}
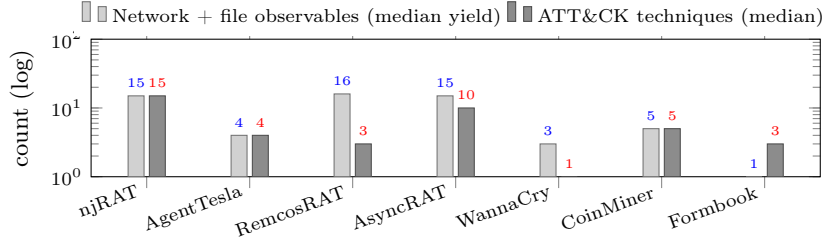

\paragraph{Worked example (AgentTesla).}
We trace one sample end-to-end. The AgentTesla sample (SHA-256 \texttt{478a9a9e\ldots}, a
32-bit Windows executable) detonates in $160$\,s and yields $16$ network and file observables
(plus $18$ registry modifications) and $12$ ATT\&CK techniques. From these the
CAPE$\rightarrow$STIX connector emits $28$ STIX objects, the $16$ observables (a
\texttt{File} carrying the sample hashes and $15$ network
\texttt{IPv4Address}/\texttt{DomainName}/\texttt{URL} objects for the contacted hosts and
the exfiltration endpoint \texttt{ftp.freshnest2u[.]com}) together with $12$
\texttt{AttackPattern}s ($16{+}12{=}28$) for techniques consistent with an information
stealer --- credentials in files (T1552.001), email collection (T1114) and application-layer
command-and-control (T1071) among them; the \texttt{Malware} object anchors these
relationships and, as in Q1, is not counted among the $28$. The confidence stage promotes the file hash to
\textsf{corroborated} once MalwareBazaar confirms the hash (recording the family tag
\emph{AgentTesla}), retains the behavioural techniques as \textsf{attested}, and marks the
single-source contacted addresses \textsf{observed}. OpenCTI merges these into one graph.
Finally the self-hosted LLM, given only this confidence-qualified evidence set, produces a
$16$-statement report of which $15$ ($93.8\%$) cite a supporting object and none reference a
non-existent one, which an analyst validates before
the intelligence is used. From submission to a validated, shareable product the critical
path is the ${\sim}160$\,s detonation; conversion, qualification, and enrichment add
sub-second overhead.

\paragraph{Stage-by-stage contribution.}
Table~\ref{tab:ablation} isolates each stage's measured contribution. The confidence model
is where the enforcement set is narrowed: of $489$ schema-valid objects only $48$ ($9.8\%$)
are enforce-eligible, keeping $90\%$ off the automated-action path; enrichment and the grounded
summary then add corroboration and auditability. This is a stage-by-stage decomposition
rather than a comparison to an external system; head-to-head baselines against raw-log LLM
summarisation and converter systems (cape2stix, plain OpenCTI) remain open.

\begin{table}[t]
\centering
\caption{Stage-by-stage contribution (35-sample corpus): each row adds one stage and reports
its measured effect. Confidence qualification narrows the enforcement set from $489$
objects to $48$ enforce-eligible.}
\label{tab:ablation}
\footnotesize
\setlength{\tabcolsep}{5pt}
\begin{tabular}{@{}lR{6.7cm}@{}}
\toprule
\textbf{Configuration} & \textbf{Measured output} \\
\midrule
CAPEv2 report only      & raw JSON; not consumable as structured CTI \\
\ +\,STIX conversion    & $489$ schema-valid objects ($100\%$), correlatable but undifferentiated \\
\ +\,Confidence model   & $48/489$ ($9.8\%$) enforce-eligible; $90\%$ routed to review or context \\
\ +\,Enrichment         & $28.6\%$ of atomic observables externally attested ($34$ MalwareBazaar consistency checks; $14$ independent ThreatFox hits) \\
\ +\,Grounded AI summary & $12.1\%$ fabricated references caught and dropped, $0$ delivered; analyst-gated \\
\bottomrule
\end{tabular}
\end{table}

\section{Related Work}
\label{sec:related}
\paragraph{CTI platforms and standards.}
Cross-border sharing and regulatory notification layers consume such
intelligence downstream~\cite{cyberbridge}.
MISP and OpenCTI are the dominant open threat-intelligence platforms, the latter natively
over STIX~2.1. \sys\ builds on OpenCTI rather than reinventing the knowledge
base, contributing the malware-analysis ingestion and the confidence layer above it, and
reuses STIX~2.1 and MITRE ATT\&CK as interchange and vocabulary; standards-based
distribution is supported by the platform but not exercised here.

\paragraph{Dynamic and evasive malware analysis.}
CAPEv2~\cite{capev2}, in the Cuckoo lineage, is the dynamic-analysis substrate \sys\
consumes; the trade-offs of dynamic analysis are surveyed by Or-Meir et
al.~\cite{ormeir2019dynamic}. Evasion is the dominant limiting factor and an
active front: hardening CAPE against VM fingerprinting~\cite{capesandbox_evasion},
symbolic-execution hybrids that recover path-dependent
behaviour~\cite{beyondsandbox}, extension to Linux and OT
architectures~\cite{samosa}, and anti-VM YARA libraries~\cite{antivm_yara}. \sys\
transforms \emph{observed} evidence, so these works bound the intelligence it can produce
and motivate its flagging of low-yield samples (Sects.~\ref{sec:eval},~\ref{sec:limits}).

\paragraph{ML-based malware representation and classification.}
A line of work turns low-level malware evidence into machine-learning representations:
binary visualisation with a convolutional network~\cite{malwaresquid} and multi-label
behavioural classification over CAPE features~\cite{decode}. \sys\ is
complementary: it does not classify malware but converts and operationalises its evidence
into shareable intelligence.

\paragraph{Malware evidence to structured CTI.}
Closest to \sys's conversion step, cape2stix~\cite{cape2stix} deterministically maps CAPE
reports to STIX~2.1 (loaded into a Neo4j graph), and AutoMAEC~\cite{automaec} standardises
dynamic traces into malware-action semantics for power-grid control systems. \sys\ does not re-implement a converter, its
connector maps CAPE$\rightarrow$STIX~2.1 directly, with coverage comparable to cape2stix
(file, network, attack-pattern), but embeds conversion in an
\emph{operational} OpenCTI pipeline that adds confidence qualification, enrichment, graph
correlation, grounded summarisation, and analyst-controlled sharing. Deterministic
CAPE$\rightarrow$STIX conversion alone is therefore no longer the novelty; the
contribution is the surrounding operational lifecycle.

\paragraph{Confidence and quality in CTI.}
Confidence is not a new concern. MISP records \emph{sightings} and decays indicators over
time~\cite{misp,misp_decay}, and STIX~2.1 carries an optional numeric \emph{confidence}
property with standard scales (Admiralty, ICD-203)~\cite{stix21}; Chen et
al.~\cite{stixgraph_ioc} score IOC quality across relevance, accuracy, timeliness,
completeness, and source confidence; EVACTI~\cite{evacti} evaluates CTI
\emph{actionability}; automatic quality assessment of structured STIX separates content
quality from provider confidence~\cite{ctiquality_venckauskas}; and weighted quality
criteria \emph{prioritise} intelligence in the MISP model~\cite{ctiquality_weighted}.
These \emph{assess} or \emph{rank} intelligence; none makes the assessment the authority
that decides whether an object may enter an automated-action path. Empirically, feeds are of uneven and often low
quality~\cite{feedquality_bouwman}, which motivates our lower-bound reading
of feed corroboration. The same concern recurs in runtime monitoring, where an
evidence-completeness status stops a gap in the event stream reading as an unqualified
all-clear~\cite{rvfabric}, and hierarchical edge-IoT monitors qualify verdicts
similarly~\cite{hybridrv,hiermon}; \sys\ makes the analogous move for CTI objects. \sys\ differs in one respect: it composes provenance, corroboration and durability into a
status that \emph{gates automated enforcement}, where prior systems record or score
confidence for an analyst to read. Its axes are a minimal
instance of those quality dimensions, and its evaluation measures STIX \emph{validity}.

\paragraph{AI-assisted CTI and its risks.}
LLMs are increasingly applied to CTI: eLLM-CTI~\cite{ellmcti} and 0-CTI~\cite{zerocti}
extract STIX from unstructured reports, ChronoCTI~\cite{chronocti} mines temporal attack
patterns, and AC\_MAPPER~\cite{acmapper} classifies content into ATT\&CK techniques.
Recent evaluations find LLMs inconsistent and overconfident on realistic
CTI~\cite{llm_unreliable_cti,llm_cti_vuln}, echoing the broader hallucination
problem~\cite{hallucination}. Complementary defences constrain what a model may \emph{do}
rather than say: closed intent vocabularies with policy rails~\cite{guardrails}, and runtime
monitors over agent tool-call traces~\cite{rvagents}. \sys\ differs on both input and safety: its LLM consumes
deterministically-converted, confidence-qualified STIX evidence rather than
natural-language reports, and every statement is either grounded in cited evidence or marked unsupported, then analyst-validated. This
restriction on the model's role is part of the contribution.

\begin{table}[t]
\centering
\caption{Where \sys\ sits relative to representative systems. ``score'' denotes an analytic
quality value an analyst reads; ``sightings'' is MISP's sighting count (MISP additionally
\emph{decays} indicators, akin to our durability axis); ``status'' a value that gates
automated action. Only \sys\ in this comparison couples per-object confidence
qualification directly to an automated-enforcement gate.}
\label{tab:related}
\footnotesize
\setlength{\tabcolsep}{4pt}
\resizebox{\columnwidth}{!}{%
\begin{tabular}{@{}lccccl@{}}
\toprule
 & \textbf{Live} & \textbf{Det.} & \textbf{Per-obj.} & \textbf{Enforce} & \\
\textbf{System} & \textbf{sandbox} & \textbf{STIX~2.1} & \textbf{confidence} & \textbf{gate} & \textbf{LLM stage} \\
\midrule
cape2stix~\cite{cape2stix}            & --         & \checkmark & --     & --         & -- \\
AutoMAEC~\cite{automaec}              & --         & MAEC       & --     & --         & -- \\
Chen et al.~\cite{stixgraph_ioc}      & --         & \checkmark & score     & --         & -- \\
EVACTI~\cite{evacti}                  & --         & \checkmark & score     & --         & -- \\
MISP~\cite{misp,misp_decay}           & --         & partial    & sightings & --         & -- \\
OpenCTI~\cite{opencti}                & --         & \checkmark & score     & --         & -- \\
eLLM-CTI~\cite{ellmcti}               & --         & LLM        & --        & --         & report$\to$STIX \\
0-CTI~\cite{zerocti}                  & --         & LLM        & --        & --         & report$\to$STIX \\
\sys                                  & \checkmark & \checkmark & status    & \checkmark & STIX$\to$narrative \\
\bottomrule
\end{tabular}}
\end{table}

\paragraph{Research gap.}
The gap is not that CTI lacks confidence, nor that sandbox output cannot be
converted: both are well served. Prior work addresses sandbox-to-STIX conversion, CTI
quality and actionability assessment, and LLM-assisted intelligence production
\emph{separately}. What is less explored, and what \sys\ provides, is the end-to-end path:
live sandbox evidence converted to structured CTI, qualified at \emph{object} level from
explicit evidence, where that qualification itself decides whether the object may enter an
automated defensive-action path. The language model sits downstream of both the
evidence and the control decision: it writes the analyst-facing narrative and determines
neither (Table~\ref{tab:related}).

\section{Discussion}
\label{sec:discussion}

\paragraph{Deterministic conversion, bounded AI.}
\sys\ separates two concerns that report-to-STIX LLM extractors~\cite{ellmcti,zerocti}
combine: the STIX \emph{representation} is produced deterministically, and the LLM only
\emph{summarises} evidence already converted and confidence-qualified. The model cannot
invent the representation, only interpret validated objects, which is why no hallucinated
reference survives the grounding check (Algorithm~\ref{alg:egmir}) though the model emitted
them on $12.1\%$ of statements. Given that LLMs are unreliable on realistic
CTI~\cite{llm_unreliable_cti,llm_cti_vuln}, this bounded role and the analyst gate are what
make the AI stage operationally acceptable; where a model drives action rather than
narrating evidence, the same discipline moves to the action interface~\cite{guardrails}.

\paragraph{Confidence as a minimal quality model.}
The three axes reuse quality dimensions studied by Chen et al.~\cite{stixgraph_ioc} and
EVACTI~\cite{evacti}; the contribution is not the axes but their composition into a total
function that gates enforcement using only signals the graph already records. A rule-based
model suits early deployment: no training labels, auditable to a SOC, and computable on
data already present. Learned calibration of the cut points is a natural extension
once operator feedback exists (Sect.~\ref{sec:limits}).

\paragraph{Coverage is the ceiling.}
The $1$--$23$ yield spread (Fig.~\ref{fig:yield}) shows that sandbox coverage, not
conversion, bounds the intelligence \sys\ produces: on low-yield samples the pipeline
surfaces unsupported content as flagged and filters fabricated references before analyst
review, so invented behaviour is never presented as fact. That is the failure mode which
keeps LLM output out of SOCs. The same sparsity surfaces upstream as a coverage signal
\sys\ reports rather than suppresses. Evasion-resistant sandboxing~\cite{capesandbox_evasion,beyondsandbox},
multi-architecture analysis~\cite{samosa}, and static anti-VM detection~\cite{antivm_yara}
would raise the floor on yield.

\paragraph{Data sovereignty.}
Running the summarisation model on operator infrastructure (Qwen2.5-14B~\cite{qwen25} via
vLLM~\cite{vllm}) keeps malware evidence on-premises. For an operator that cannot send samples or derived
evidence to a third-party cloud, this is a precondition for using an LLM at all.

\section{Limitations and Threats to Validity}
\label{sec:limits}
\emph{Deployment scope.} \sys\ is a project prototype evaluated against a single live
CAPEv2 deployment, not yet an operational installation inside a national-CSIRT SOC, so
absolute figures are indicative and the confidence and grounding behaviour await
validation in production SOC/CSIRT use. \emph{Corpus.} The evaluation corpus is detonated through the live sandbox;
family labels come from the source repository and the ATT\&CK figures are \emph{recovered
coverage} from the sandbox's own technique mapping, not an independent detection-accuracy
measurement. \emph{Untuned thresholds.} The confidence cut points of
Sect.~\ref{sec:confidence} are design defaults. \emph{Read layer.} The dashboard
service is pilot-grade: it currently reads indicators by string-parsing STIX patterns and
labels rather than querying structured observables, and lacks authentication. \emph{Sanitisation.} The assessment tier normalises findings but does not
\emph{sanitise} them: web-scanner findings carry internal URLs and code findings source
fragments, so such objects need redaction and a marking policy before leaving the operator. \emph{Enrichment scope.}
Enrichment is limited to two open-feed lookups (passive DNS, network-ownership,
geolocation and reputation sources are not consulted).
\emph{AI summaries.} Q7 measures citation \emph{validity} for grounded statements, whether
cited evidence identifiers resolve to real objects, not semantic faithfulness; the light
analyst check found no unsupported claims among the grounded statements but is not a full
inter-rater study. The
$12.1\%$ dropped-reference rate shows the check removes fabricated citations.
\emph{Evasion.}
Dynamic analysis is defeated by anti-VM and anti-analysis malware and misses dormant code
paths; the confidence model mitigates this by preferring corroborated and durable
evidence but cannot recover behaviour a sample never exhibited. \emph{Single detonation.}
Each sample is detonated once under one execution profile, and yield is network- and
time-dependent.
\emph{Feed trust.} The $k{\ge}1$ auto-enforce gate trusts open feeds; a
poisoned entry or a benign indicator listed in ThreatFox could promote a benign observable,
cross-sensor corroboration is more robust than a single feed hit. \emph{Enforcement.} \sys\
represents the enforcement decision as an object status; wiring it to an executed control (a
blocklist push to detection and response tooling) is unevaluated, so the loop is demonstrated
up to the gate, not through to an executed action. \emph{Ethics and availability.} All
malware was detonated in an isolated, CYBERGUARD-controlled CAPEv2 sandbox. Network routing was
enabled so samples exhibited realistic behaviour; egress was sandbox-confined with no lateral
access to project or production systems, and the only external contacts were the samples'
own command-and-control callbacks (reflected in the ThreatFox corroboration), not traffic
directed at third parties or victims. The connectors, confidence model, and evaluation
scripts will be released for replication.

\section{Conclusion}
\label{sec:conclusion}
\sys\ turns the detailed but fragmented output of a malware sandbox into
actionable intelligence for critical-infrastructure SOCs and national CSIRTs. It
converts live CAPEv2 analyses to STIX~2.1 and correlates them in an OpenCTI knowledge
graph; it attaches an explicit, evidence-derived confidence status to every object from
provenance, cross-source corroboration, and observation durability, strengthened by
lightweight open-feed corroboration; and it generates an evidence-grounded malware
summary in which each AI statement is either traceable to the STIX evidence that supports
it or explicitly marked unsupported, while fabricated evidence references are removed
before analyst validation. Implemented within the CYBERGUARD project and
evaluated on a labelled malware corpus detonated through a live CAPEv2 sandbox, \sys\
shows that coupling machine-generated sandbox evidence to explicit confidence and auditable AI turns an
undifferentiated indicator stream into corroborated, durability-ranked, and explainable
intelligence, an operational layer between malware analysis and actionable
critical-infrastructure defence.

\paragraph{Future work.} Four directions follow from Sect.~\ref{sec:limits}. \emph{External
validity:} replicate across multiple CAPEv2 deployments and detonation profiles, separating
pipeline behaviour from one sandbox's characteristics. \emph{Closing the enforcement loop:}
execute and measure the response actions the gate authorises, and tune the default
thresholds against operator feedback. \emph{Corroboration breadth and feed trust:} widen
enrichment beyond two open feeds while modelling the trust the gate places in them, since an
eligibility decision inherits its sources' weaknesses. \emph{Semantic grounding:} extend the
citation-validity check to a faithfulness study against analyst judgement, including a
head-to-head against raw-log summarisation. Behavioural abstraction, multi-sample campaign
narratives, and a production-grade analyst interface remain open.

\begin{credits}
\subsubsection{\ackname}
This work has received funding from the European Union's Digital Europe Programme under
grant agreement No~101190251 (CYBERGUARD).

\subsubsection{\discintname}
The authors have no competing interests to declare that are relevant to the content of
this article.
\end{credits}

\bibliographystyle{splncs04}
\bibliography{references}

\end{document}